\documentclass[letterpaper,review]{elsarticle}
\makeatletter
\def\ps@pprintTitle{%
 \let\@oddhead\@empty
 \let\@evenhead\@empty
 \def\@oddfoot{\centerline{\thepage}}%
 \let\@evenfoot\@oddfoot}
\makeatother

\usepackage{natbib}
\usepackage{cancel}
\usepackage{pifont} 
\usepackage{amsthm}         
\usepackage{amsmath} 	   
\usepackage{amssymb}       
\usepackage{amsfonts}
\usepackage{graphicx} 	    
\usepackage{verbatim}
\usepackage{epsfig,bbm}
\usepackage{color}
\usepackage{colortbl}
\usepackage{float}
\usepackage{multirow}

\usepackage{subcaption}

\definecolor{babyblue}{rgb}{0.54, 0.81, 0.94}
\definecolor{corn}{rgb}{0.98, 0.93, 0.36}

\begin{document}

\begin{frontmatter}

\title{Rapidly Descending Dark Energy and the End of Cosmic Expansion}

\author[add1]{Cosmin Andrei}
\author[add2]{Anna Ijjas }
\author[add1]{Paul J. Steinhardt\corref{cor1}}
\ead{steinh@princeton.edu}
\cortext[cor1]{Corresponding author}

\address[add1]{Department of Physics, Princeton University, Princeton, NJ 08544, USA}
\address[add2]{Center for Cosmology and Particle Physics, Department of Physics, New York University, New York, NY 10003 USA}

\begin{abstract}
If dark energy is a form of quintessence driven by a scalar field $\phi$ evolving down a monotonically decreasing potential $V(\phi)$ that passes sufficiently below zero, the universe is destined to undergo a series of smooth transitions.  The currently observed accelerated expansion will cease; soon thereafter, expansion will come to end altogether; and the universe will pass into a phase of slow contraction.   In this paper, we consider how short the remaining period of expansion can be given current observational constraints on dark energy.  We also discuss how this scenario fits naturally with cyclic cosmologies and recent conjectures about quantum gravity.
\end{abstract}

%\begin{keyword}
%quintessence, dark energy, supernovae, cyclic universe, quantum gravity
%\end{keyword}

\end{frontmatter}

{\it Introduction.} 
In the $\Lambda$CDM model, dark energy takes the form of a positive cosmological constant, in which case the current period of accelerated expansion will endure indefinitely into the future \cite{Turner:2018bcg}.  An alternative is that  the current vacuum is metastable and  has positive energy density.   If it is separated by an energy barrier from a true vacuum phase with zero or negative vacuum density, then accelerated expansion will be ended by the nucleation of a bubble of true vacuum that grows to encompass us. Until that moment, cosmological observations will be indistinguishable from  the $\Lambda$CDM picture.  Without extreme fine tuning, the time scale before a bubble will nucleate \cite{Coleman:1977py} and pass our location can be exponentially many  Hubble times in the future (see, for example, Refs.~\cite{Degrassi:2012ry,Branchina:2014rva}).  (Here and throughout, `Hubble time' refers to $H_0^{-1} \approx 14$~Gy where $H_0$ is the current Hubble expansion rate.) Also, the ultra-relativistic bubble wall will likely destroy all observers in its path, so there will be no surviving witnesses to the end of accelerated expansion \cite{Coleman:1977py}.

A third possibility, to be considered here, is that the dark energy is a type of quintessence due to a scalar field $\phi$ evolving down a monotonically decreasing potential $V(\phi)$ \cite{Caldwell:1997ii}.     Since the current value of $V(\phi_0)$ is extraordinarily small today as measured in Planck mass units, there is a wide range of forms for $V(\phi)$ that pass through zero and continue to large negative values where $V(\phi) \ll -V(\phi_0)$.  In this case, the equations of motion of Einstein's general theory of relativity dictate that the universe is destined to undergo a remarkable series of smooth transitions  \cite{Alam:2003rw,Steinhardt:2001st,Ijjas:2018qbo}.   

First, as the positive potential energy density decreases and the kinetic energy density comes to exceed it, the current phase of accelerated expansion will end and smoothly transition to a period of decelerated expansion. Next, as the scalar field continues to evolve down the potential, the potential energy density will become sufficiently negative that the total energy density ($\propto H^2(t)$) and, consequently, the Hubble parameter $H(t)$, will reach zero.  Consequently, expansion ($H>0$) will stop altogether and smoothly change to contraction ($H<0$).  More  precisely, the transition will be to a phase of {\it slow contraction} \cite{Steinhardt:2001st,Ijjas:2018qbo}  in which the Friedmann-Robertson-Walker (FRW) scale factor $a(t) \propto |H^{-1}|^{\alpha}$  where $\alpha < 1/3$.

In this paper, we consider how soon these transitions could begin.  That is, what is the minimal time, beginning from the present  ($t=t_0$), before expansion ends and contraction begins given current observational constraints on dark energy and without introducing extreme fine-tuning?   One might imagine the answer is one or more Hubble times given how well $\Lambda$CDM is claimed to fit current cosmological data.   

\vspace{.1in}
\noindent
{\it The Q-SC-CDM model.}
We introduce the acronym Q-SC-CDM to refer to cold dark matter (CDM) models with a phase of quintessence-driven (Q)
accelerated expansion transitioning in the future to decelerated expansion and subsequently to slow contraction (SC), where all phases are dominated by a scalar field $\phi({\bf x}, t)$ evolving down a potential $V(\phi)$.

The series  of continuous transitions can be understood by tracking the 
the  total cosmic equation-of-state $\varepsilon_{\rm TOT}(t)$, including both matter and dark energy densities:
\begin{eqnarray}
\label{epsTOT} 
\varepsilon_{\rm TOT} & \equiv  & \frac{3}{2}\left(1+ \frac{p_{\rm TOT}}{\rho_{\rm TOT}}\right)  \\
 & \equiv  & 3 \left( \frac{\frac{1}{2} \dot{\phi}^2 +\ \frac{1}{2} \frac{\rho_{\rm m}^0}{a^3}} {\frac{1}{2} \dot{\phi}^2+V+\frac{\rho_{\rm m}^0}{a^3}}\right),
\end{eqnarray}
where $p_{\rm TOT}$ and $\rho_{\rm TOT}$ are the total pressure and energy density, respectively; $p_{\rm Q}\equiv   \frac{1}{2} \dot{\phi}^2-V$  is the scalar field pressure; $\rho_{\rm Q}  \equiv   \frac{1}{2} \dot{\phi}^2+V$ is the scalar field energy density; and $\rho_{\rm m}^0$ is the current (pressureless) matter density.   (Dot represents the derivative with respect to FRW time.)

As $V(\phi)$ approaches zero from above, $\varepsilon_{\rm TOT}$
grows to be greater than  one, which marks the end of accelerated expansion ($\ddot{a}> 0$) and the beginning of decelerated expansion ($\ddot{a}<0$) according the Friedmann equation: 
\begin{equation} 
\label{Facc}
 \frac{\ddot{a}}{a} = \frac{4 \pi G }{3} (1- \varepsilon_{\rm TOT})  \, \rho_{\rm TOT},
 \end{equation} 
 where $G$ is Newton's constant. 
The value of $\varepsilon_{\rm TOT}$ continues to rise as $V(\phi)$ passes below zero until $V(\phi)$  becomes sufficiently negative that $\rho_{\rm TOT}$ reaches zero.  According to the Friedmann constraint,
\begin{equation}
\label{FHam}
H^2  = \left(\frac{\dot{a}}{a}\right)^2 = \frac{8 \pi G}{3} \rho_{\rm TOT},
\end{equation}
the expansion rate
$H(t)$ also reaches zero at that point.  (Spatial curvature is negligible today and throughout these stages.)   The Friedmann equations above combined with the equation-of-motion for the scalar field
\begin{equation}
\label{eqphi}
\ddot{\phi} + 3 H \dot{\phi}= -V_{,\phi}
\end{equation}  
 dictate that the field continues to evolve down its potential and that $H$ continues to decrease.  This means that 
$H$ passes through zero; {\it i.e., } the universe necessarily begins to contract.  (Note that Eq.~(\ref{FHam})  ensures that $\rho_{\rm TOT}$ cannot become negative; rather $\rho_{\rm TOT}$    increases from zero once contraction begins.)  For the steep potentials of interest in this paper that minimize the time until expansion ends, $\varepsilon_{\rm TOT}$ becomes $\gg 3$, corresponding to 
$a(t)\propto |H^{-1}|^{\alpha}$ where $\alpha \approx 1/\varepsilon_Q \ll 3$, the condition for slow contraction.

%FIGURE1
\begin{figure}[tb]
\begin{center}
\includegraphics[width=3.3in,angle=-0]{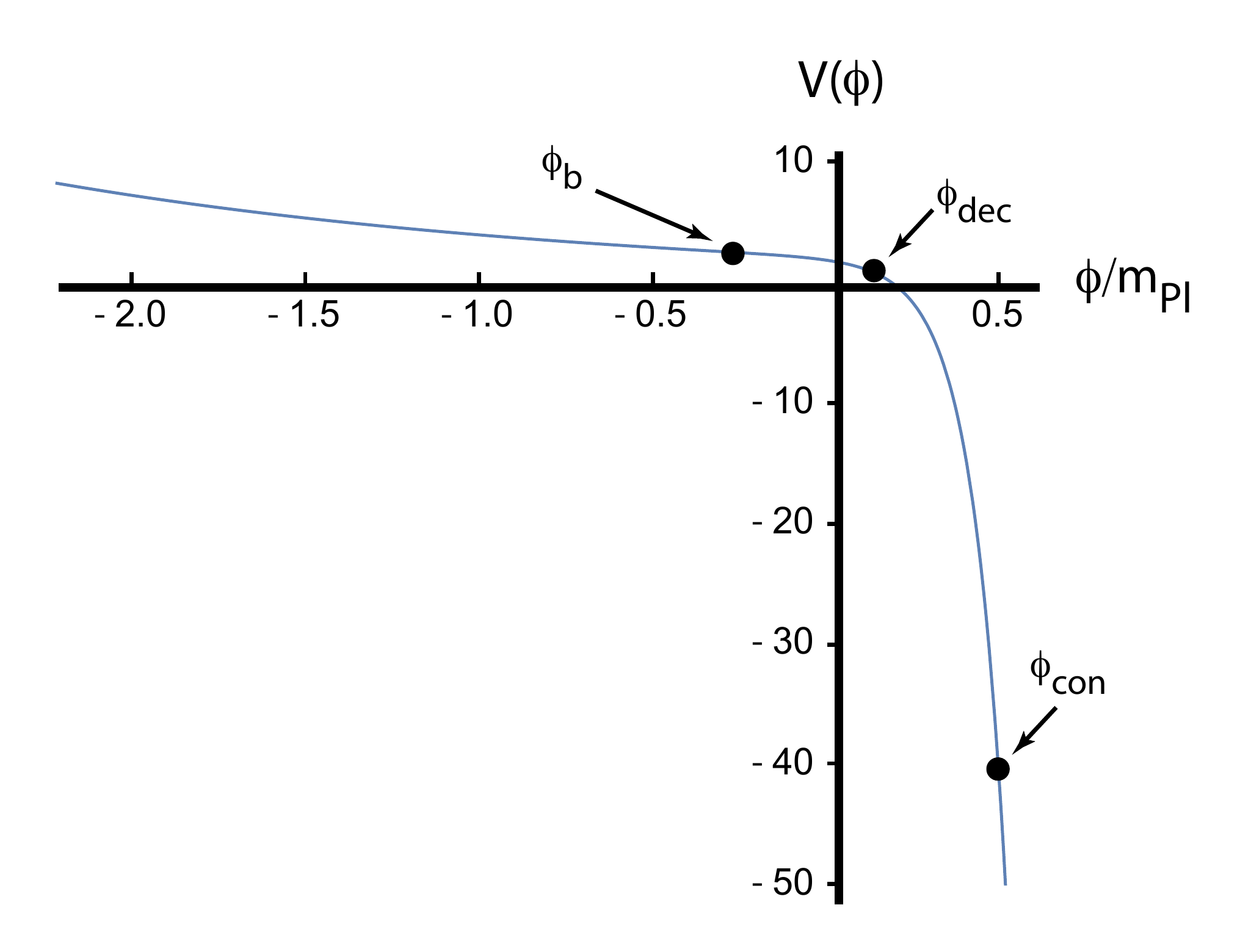}
\end{center}
\caption{The Q-SC-CDM scalar field potential in Eq.~(\ref{eq1}),  $V(\phi)$ (in units of $H_0^2 m_{\rm Pl}^2$) vs. $\phi/m_{\rm Pl}$ with $M= 1.7  m_{\rm Pl}$ and $m =0.1 m_{\rm Pl}$.  As described in the text, the field is fixed by Hubble friction near $\phi_b$ until around redshift $z=3$ ($t=t_b = -0.8 H0^{-1}$); it then evolves to $\phi=0$ today ($t=t_0$); continues to evolve to $\phi_{\rm dec}>0$, at which time ($t_{\rm dec}$) accelerated expansion turns to decelerated expansion; and then $\phi$ evolves further until $V(\phi)$ becomes sufficiently negative (at $t=t_{\rm con}$) that the Hubble parameter $H$ passes through zero, the expansion phase ends and slow contraction begins.} 
\label{Fig1}
\end{figure} 
%%%

Notably, in contrast to the case of quantum tunneling from a metastable phase, the entire sequence of transitions from accelerated expansion to slow contraction would be sufficiently smooth and slow that observers could safely survive and  bearing witness to each stage.

To determine the minimal time before these transitions could occur, we consider Q-SC-CDM potentials of the form
\begin{equation}
\label{eq1}
V(\phi) = V_0 e^{-\phi/M} -  V_1 e^{\phi/m}.
\end{equation}
The initial conditions and parameters  $V_{0,1}>0$  are chosen such that  $\phi$ evolves from $\phi 
\le 0$ in the past ($t \le t_0$) to $\phi>0$ in the future, as illustrated in Fig.~\ref{Fig1}.  The first (positive) potential term dominates during the quintessence-driven accelerated expansion phase (which includes the past and part of the future; and the second (negative term) dominates beginning at some time in the future.

The initial value of the scalar field at the beginning of the dark energy dominated phase, $\phi=\phi_b$, is uniquely determined once the parameters are chosen such that $\Omega_{\rm m}^0$ and $\Omega_{\rm DE}^0$,  the ratios of the present dark energy and matter densities to the critical density, are in accord with current observational limits.  More precisely, extrapolating the Friedmann equations and the equation of motion for $\phi$  back in time beginning from $\phi_0=0$, one finds that the scalar field becomes frozen  by Hubble friction (the $3H \dot{\phi}$ term in Eq.~(\ref{eqphi})) as matter dominates over dark energy, which is what sets the value of $\phi_b$.

 %FIGURE2
\begin{figure}[tb]
\begin{center}
\includegraphics[width=3.3in,angle=-0]{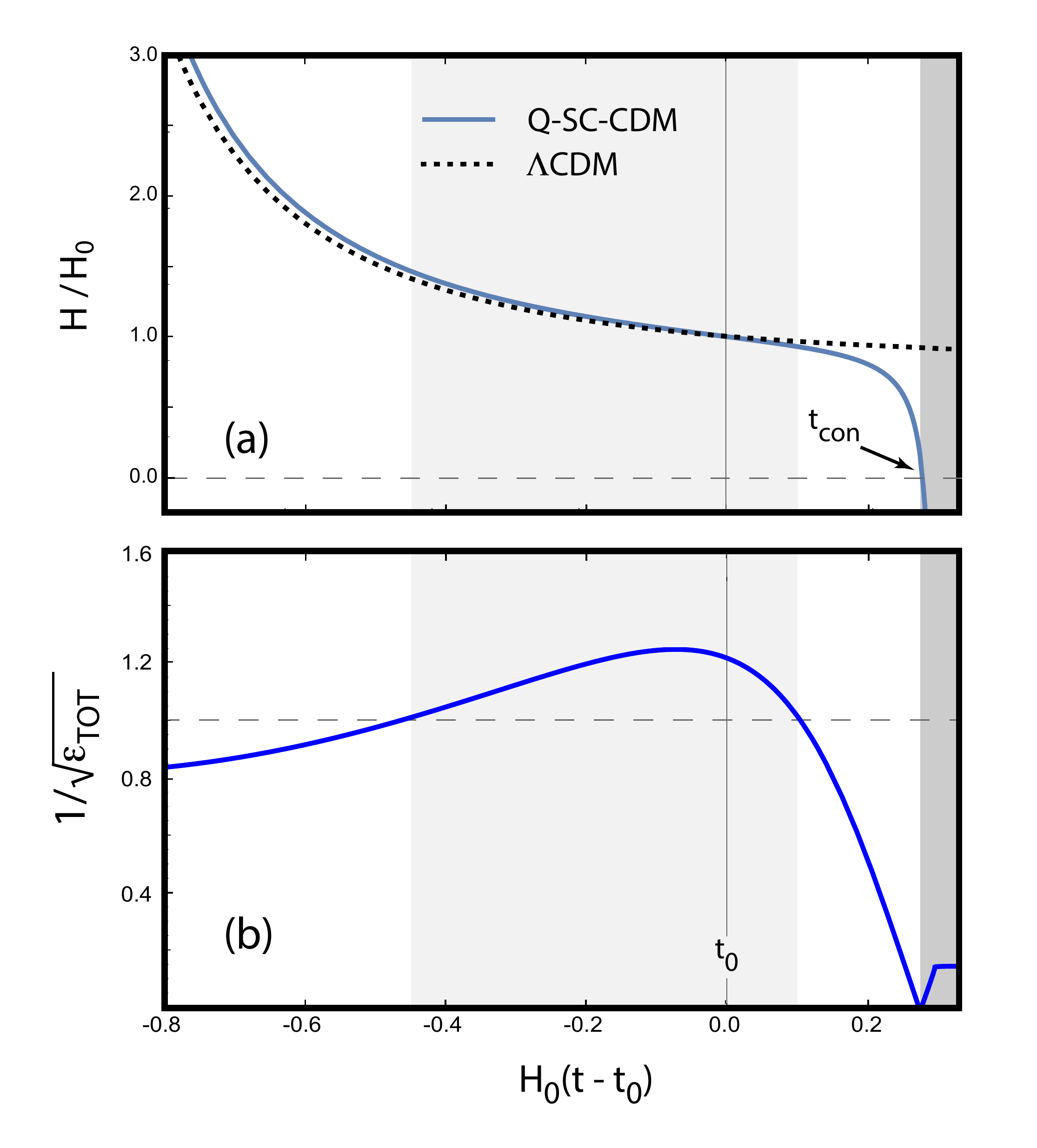}
\end{center}
\caption{(a) The Hubble parameter $H(t)$ for the best-fit $\Lambda$CDM model (dotted curve) and Q-SC-CDM (solid curve) models. (b) A plot of  $1/\sqrt{\varepsilon_{\rm TOT}}$ vs. $t$ for the Q-SC-CDM model depicted in Figs.~\ref{Fig1} and~Fig.~\ref{Fig2}a.   Unshaded regions are periods of decelerated expansion.  The light grey shaded region  is the phase of accelerated expansion  ($H>0$ and $\varepsilon_{\rm TOT}  <1$ beginning about redshift $z=0.75$).  The dark shaded region is the phase slow contraction ($H<0$ and $\varepsilon_{\rm TOT}  >3$) that begins at $t=t_{\rm con}$.} 
\label{Fig2}
\end{figure} 
%%% 

The shortest time before the end of expansion will occur for the steepest potential  ({\it i.e.}, the largest allowed  value of $|V_{,\phi}/V|$ or smallest values of $M$ and $m$ in Eq.~(\ref{eq1}).   As shown in Ref.~\cite{Agrawal:2018own}, a positive exponential potential with $M \approx 1.7 m_{\rm Pl}$,  is the steepest potential compatible (to within $2 \sigma$) with current observations, where $m_{\rm Pl} = 1/\sqrt{8 \pi G}$ is the reduced Planck mass.  The negative potential term is negligible in the past, so $m$ is not constrained by observations.  However, we also want to avoid extreme fine-tuning.  The ratio $m/M$ can be viewed as the figure-of-merit for judging fine-tuning; we therefore confine our study to values of $10^{-2} < m/M  < 1$, though our results below can be used to infer the consequences for a narrower or wider range.

\vspace{.1in}
\noindent
{\it A worked example.}
Fig.~\ref{Fig1} illustrates the case where $m= 0.1 m_{\rm Pl}$. The potential parameters $V_0=2.1 H_0^2 m_{\rm Pl}^2$ and $V_1=0.28 H_0^2 m_{\rm Pl}^2$
are chosen such that $\Omega_{\rm m}^0=0.29$ and $\Omega_{\rm DE}^0=0.71$, within current observational limits  \cite{Planck:2018vyg,ACT:2020gnv}.

For the negative exponential potential term in Eq.~(\ref{eq1})  that dominates by the time $H$ reaches zero and contraction begins, there is an attractor solution with $a(t) \propto |H^{-1}|^{\alpha}$ where 
 $\alpha=2 (m/m_{\rm Pl})^2$; for the worked example with $m=0.1 m_{\rm Pl}$, $\alpha = 0.02 \ll 1/3$, the signature of slow contraction.

Fig.~\ref{Fig2}a compares the evolution of  of the Hubble parameter $H(t)$ as a function of time (in units of $H_0^{-1}$) for the best-fit  $\Lambda$CDM  model \cite{Planck:2018vyg,ACT:2020gnv} and the best-fit Q-SC-CDM model with $m=0.1 m_{\rm Pl}$.   In the past (negative values of $t$), the two curves are nearly parallel; the first term in Eq.~\ref{eq1} dominates; and the expansion rate is accelerating.  The two curves diverge going forward in time.  Accelerated expansion occurs forever in the $\Lambda$CDM model but ends and eventually transitions to contraction (at $t=t_{\rm con}$ where $H$ passes through zero) in the   Q-SC-CDM model.

%FIGURE3
\begin{figure}[tb]
\begin{center}
\includegraphics[width=3.3 in,angle=-0]{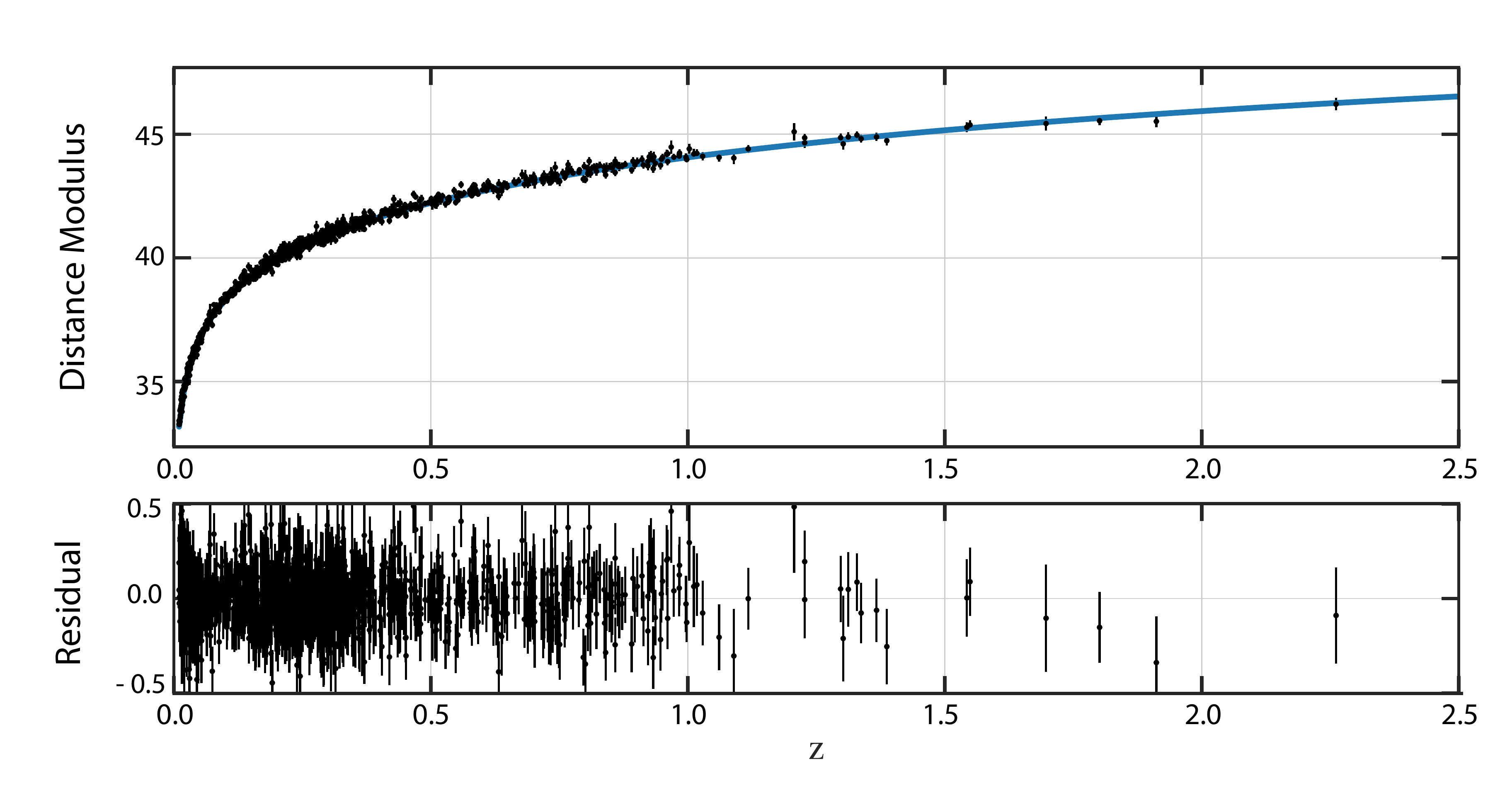}
\end{center}
\caption{The predicted luminosity-redshift relations (distance modulus vs. $z$) for the Q-SC-CDM model shown in preceding figures compared to the Hostz-only supernovae catalog compiled in Ref.~\cite{Steinhardt:2020kul} from the Pantheon dataset \cite{Pan-STARRS1:2017jku}, a fit to within $2 \sigma$. } 
\label{Fig3}
\end{figure} 
%%%

The evolution of the  total cosmic equation-of-state $\varepsilon_{\rm TOT}(t)$ 
is shown in Fig.~\ref{Fig2}b.    According to the Friedmann equation~\eqref{Facc}, $\ddot{a} \propto (1-
\varepsilon_{\rm TOT} )$, so, when $H>0$,  acceleration corresponds to $\varepsilon_{\rm TOT} <1$ (light shade region in the middle) and deceleration corresponds to $\varepsilon_{\rm TOT} >1$ (unshaded regions).    
From the figure, one can observe when quintessence first dominates sufficiently for accelerated expansion to begin and when the accelerated expansion ends in the future.  During contraction ($H<0$),  $\varepsilon_{\rm TOT} 
\gg 3$, corresponding to slow contraction (dark shaded region).  The value of $\varepsilon_{\rm TOT}$ rapidly asymptotes to $\frac{1}{2}(m_{\rm Pl}^2/ m^2) = 50 \gg 3$ after contraction begins.

Fig.~\ref{Fig3} shows the  predicted luminosity-redshift relation for the Q-SC-CDM model compared to current supernovae observations \cite{Steinhardt:2020kul}, demonstrating that the goodness of fit is $2 \sigma$.  The distance modulus is the conventional way of parameterizing the apparent luminosity of an object at redshift $z$; for standard candles, the modulus is  equal to $5\, { \rm log}_{10} [d_p(z)(1+z)/(10 \, {\rm pc})]$ where $d_p(z)$ is the luminosity distance.
 (Because the evolution of $H(t)$ is so similar in the past to the $\Lambda$CDM model, the Q-SC-CDM model fits no better or worse; so it does not alleviate or exacerbate the current `$H_0$ problem'  \cite{Freedman:2021ahq}.)

\begin{figure}[tb]
\begin{center}
\includegraphics[width=3.3 in,angle=-0]{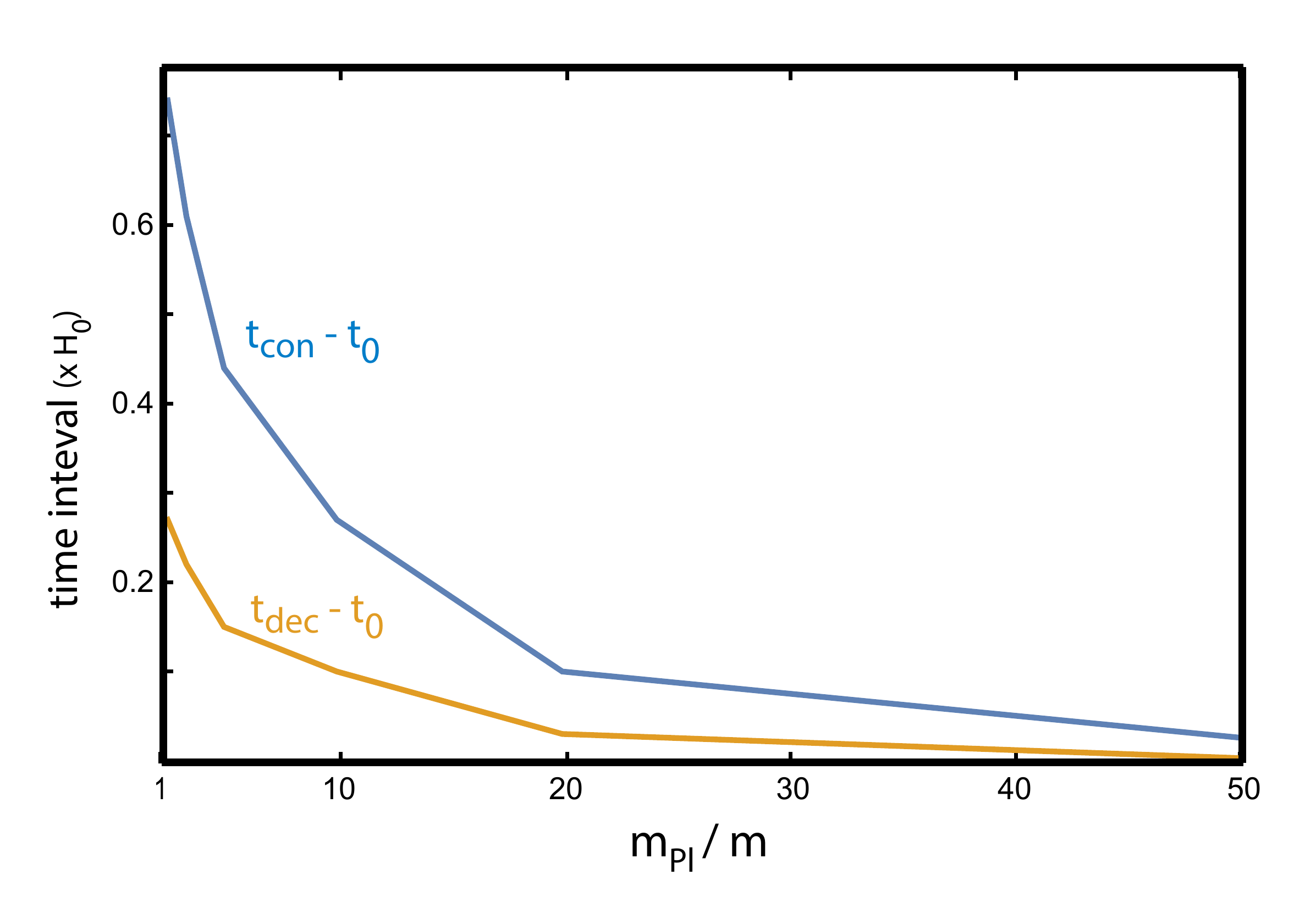}
\end{center}
\caption{The minimum time intervals (in units of Hubble time) between now and the end of accelerated expansion 
($t_{\rm dec}-t_0$; lower curve) and between the end of expansion ($t_{\rm con}-t_0$; upper curve)  as a function of $m_{\rm Pl}/m$. 
} 
\label{Fig4}
\end{figure}

\vspace{.1in}
\noindent
{\it Results and Discussion.} 
 For the worked example in  Fig.~\ref{Fig2}b with  $m_{\rm Pl}/m =10$, $t_{\rm dec} -t_0=  0.1 H_0^{-1}$ and $t_{\rm con}-t_0= 0.27 H_0^{-1}$ --  less than a Hubble time and on the order of billions of years.  For steeper potentials ($m_{\rm Pl}/m > 10$), the minimal times are predicted to be even sooner, as shown in  
Fig.~\ref{Fig4}.  For each value of $m$, the potential parameters are chosen to ensure fits to current observational constraints on $\varepsilon_{\rm TOT} $, $\Omega_m^0$ and $\Omega_{DE}^0$ at the 2$\sigma$ level.   

  %FIGURE4
%%%     co
 
These minimum time intervals before the end of acceleration and the end of expansion are each strikingly soon, cosmologically speaking.  In fact, they can be compared to geologic timescales.  For $m_{\rm Pl}/m =10$, the minimum time remaining  before the end of expansion, for example, is roughly equal to the period since life has existed on Earth; for $m_{\rm Pl}/m =50$, the time interval remaining before the end of acceleration is less than the time since the Chicxulub asteroid brought an end to the dinosaurs.
 
 Yet, curiously, we could not detect the oncoming dramatic cosmic events given the best-available observations today.   The problem is that accurate cosmological measures of the expansion rate and other cosmological parameters are based on observations of the cosmic microwave background, baryon acoustic oscillations and distant objects, like supernovae, whose detected light was emitted far in the past, whereas, as we have shown, the transitions to deceleration and slow contraction may all occur within a small fraction of a Hubble time.  For this reason, it is a challenge to detect the end of contraction even when the time is nigh.  Improvements in measuring $\varepsilon_{\rm TOT}$ and especially its time variation would be a generic approach.  In the example in Fig.~\ref{Fig2}, for example, the prediction is that $1/\sqrt{\varepsilon_{\rm TOT}}$ has already reached a maximum and is beginning to decrease because the $\dot{\phi}$ has begun to increase significantly.  As $\dot{\phi}$ increases, there may be a rich set of additional observable effects to pursue, depending on the couplings of $\phi$ to other fields \cite{Berghaus:2020ekh,Agrawal:2019dlm,Bloch:2019bvc,Csaki:2020zqz}.  These possibilities and other potential observable consequences will be explored in future work.

What happens after the transition from expansion to contraction is model dependent.
Notably, Q-SC-CDM  connects naturally with the recently proposed  cyclic  model of the universe  \cite{Ijjas:2019pyf} in which the big bang is replaced by a non-singular classical bounce that connects a previous period of slow contraction to a subsequent period of radiation-, matter- and dark energy-dominated expansion.   Slow contraction is an essential element because it is responsible for making the universe homogeneous, isotropic and spatially flat and for setting the background conditions needed to generate a nearly scale invariant spectrum of density perturbations \cite{Cook:2020oaj}.  
By construction, a cyclic model demands that each dark energy phase, including the current one,  comes to an end and transitions smoothly to the next phase of slow contraction in order to set the large-scale properties of the universe for the cycle to come.  The slow contraction phase endures for a period of order a billion years before the universe transitions to a new phase of expansion and reheats to temperatures well above the electroweak scale ($10^{15}$~K) that would likely vaporize all pre-existing matter other than black holes. 

Q-SC-CDM provides all the necessary conditions and can easily be incorporated as part of each cycle; see, for example, Ref.~\cite{Ijjas:2019pyf}.  In this case,
 there is an interesting connection between the results presented here and the `why now?' problem.  The `why now' problem is the mystery of explaining why dark energy only began to dominate recently, just as the galaxies like our Milky Way formed and planets capable of supporting life first evolved.  
  In a cyclic universe with a very short time between now and the end of expansion, the period of galaxy formation and the onset of accelerated expansion is the longest interval of time and comprises the largest spatial volume, which makes the `why now?' issue seem less mysterious \cite{Page:2006dt,Bedroya:2019snp}.

Q-SC-CDM also dovetails with recent conjectures about quantum gravity constraints on dark energy.   The so-called `swampland conjectures' \cite{Ooguri:2006in,Palti:2019pca,Agrawal:2018own,Bedroya:2019snp}  rule out  the possibility that dark energy is a cosmological constant or that the  energy density is associated with a metastable phase,  and only allow the possibility considered here -- that dark energy is due to a quintessence field with a monotonically decreasing $V(\phi)$.   It does not require that $V(\phi)$ pass below zero, but it is allowed and occurs for a wide range of parameters. 
The swampland conjectures also place a quantitative constraint on how long the current period of accelerated expansion might last based  on the condition that ultraviolet sub-Planckian fluctuations should never expand to scales larger than the Hubble radius \cite{Bedroya:2019snp}.   The {\it upper bound} to the end of expansion  according to these conjectures is about 2.4 trillion years or about 160 Hubble times.  The {\it lower bound} derived  here based on independent reasoning is consistent.

Curiously, three different theoretical developments point to the same outcome: the end of expansion could occur surprisingly soon.

 \vspace{0.1in}
\noindent
{\it Acknowledgements.} 
We thank  A. Sneppen and C.L. Steinhardt for providing the supernova plot in Fig. 3  and  for useful discussions.  We also thank A. Bedroya and P. Agrawal for helpful comments.
The work of A.I. is supported by  the Simons Foundation grant number 663083.
P.J.S. is supported in part by the DOE grant number DEFG02-91ER40671 and by the Simons Foundation grant number 654561.  

\bibliographystyle{apsrev}
\bibliography{EndofExpansion}

\end{document}